\documentclass[10pt,fleqn]{revtex4}
\usepackage{graphicx}
\usepackage{float}
\usepackage{textcomp}
\usepackage{color}
\usepackage{amsmath}
\usepackage{psfrag,overpic}
\usepackage{ulem}
\usepackage{enumitem}
\usepackage{siunitx}

\begin{document}

\title{Full Vectorial Modeling of Second Harmonic Generation in III-V-on-insulator Nanowires}
\author{Charles~Ciret,$^{1}$ Koen~Alexander,$^{2,3}$ Nicolas~Poulvellarie,$^{2,3,4}$ Maximilien~Billet,$^{2,3,4}$Carlos~Mas~Arabi,$^{4}$ Bart~Kuyken,$^{2,3}$ Simon-Pierre~Gorza,$^4$ and Fran\c cois~Leo$^{4,*}$}

\affiliation{$^1$ Laboratoire de Photonique d'Angers EA 4464, Université d'Angers , Angers, France}
\affiliation{$^2$ Photonics Research Group, Ghent University-IMEC, Ghent, Belgium}
\affiliation{$^3$ Center for Nano- and Biophotonics (NB-Photonics), Ghent University, Ghent, Belgium}
\affiliation{$^4$ OPERA-Photonique, Universit\'e libre de Bruxelles, Brussels, Belgium}
\email{francois.leo@ulb.ac.be}

\begin{abstract}
We model second harmonic generation in subwavelength III-V-on-insulator waveguides. The large index contrast induces strong longitudinal electric field components that play an important role in the nonlinear conversion. We show that many different waveguide dimensions are suitable for efficient conversion of a fundamental quasi-TE pump mode around the 1550 nm telecommunication wavelength to a higher-order second harmonic mode.

\end{abstract}
\maketitle

\section{Introduction}

The first demonstration of second harmonic generation (SHG) in 1961 paved the way for decades of research on nonlinear interactions in dielectrics~\cite{franken_generation_1961}. The quest for efficient conversion is still relevant today as key regions of the electromagnetic spectrum lack suitable laser sources. Other applications such as squeezed light generation~\cite{paschotta_bright_1994} and frequency comb stabilization~\cite{holwarth_optical_2000} would also benefit from efficient frequency converters.

The advent of integrated photonic platforms the last decade revolutionized frequency conversion.
The large nonlinear coefficients as well as the high index contrast inherent to integrated photonics allows for strong nonlinear interaction at low power. Many instances of integrated second harmonics generation have been reported, with novel, low-loss, LiNbO$_3$ on insulator and III-V-on insulator platforms currently holding the record normalized conversion efficiency~\cite{wang_ultrahigh_2018,chang_heterogeneously_2018}.
In most theoretical analysis, the light is approximated by a purely transverse mode such that a single incoming polarization state and spatial profile is considered. In practice however, more complex nonlinear wave mixing can be expected because the optical modes that propagate unperturbed in high index contrast waveguides display large longitudinal components. Importantly, the spatial distribution of the longitudinal component of the electric field is very different from that of its transverse counterpart. 
While many full-vectorial analysis of nonlinear coupling in nanowaveguides have been reported~\cite{kolesik_nonlinear_2004,chen_theory_2006,koos_nonlinear_2007,vahid_full_2009,daniel_vectorial_2010,alloatti_second_2012,alexander_electrically_2017}, to the best of our knowledge, second harmonic generation is yet to be studied.
Consequently we here derive the ordinary differential equations describing the nonlinear coupling of a fundamental mode at $\omega_0$ to its second harmonic at $2\omega_0$ in a quadratically nonlinear waveguide. We predict ultra-efficient conversion for a wide range of waveguide dimensions and study the impact of the propagation direction.
We focus on III-V semiconductor wire waveguides as this article serves as theoretical support for our recent experimental results~\cite{poulvellarie_second_2019} but our analysis can be adapted to study other platforms.

\section{General framework}

\subsection{Linear waveguides}

We start by discussing the properties of bound modes in a nonabsorbing linear waveguide~\cite{snyder_optical_1983}. We consider a III-V-insulator wire waveguide as shown in Fig.~\ref{figcrossing}. An electromagnetic wave oscillating at $\omega_j$ propagating in the waveguide must satisfy the source-free Maxwell equations: 
\begin{eqnarray}
\nabla \times \mathbf{\tilde{E}}_{0}(\mathbf{r},\omega_j)&=& i\omega_j \mu_0 \mathbf{\tilde{H}}_{0}(\mathbf{r},\omega_j),\label{eq:MaxwellCurlE}\\
\nabla \times \mathbf{\tilde{H}}_{0}(\mathbf{r},\omega_j)&=& -i\omega_j \epsilon_0n^2 \mathbf{\tilde{E}}_{0}(\mathbf{r},\omega_j),\label{eq:MaxwellCurlH}\\
\nabla\cdot[n^2\mathbf{\tilde{E}}_{0}(\mathbf{r},\omega_j)]&=& 0,\\
\nabla\cdot\mathbf{\tilde{H}}_{0}(\mathbf{r},\omega_j)&=& 0,
\end{eqnarray}
where $n(\mathbf{r}_\perp)$ is the index of the unperturbed waveguide cross-section. 
The translational invariance allows to write the guided mode as a spatial distribution of the electric and magnetic field with a fixed propagation constant. They read:
\begin{eqnarray}
\mathbf{\tilde{E}}_{0}(\mathbf{r},\omega_j)&=& a_0\frac{\mathbf{e}_j(\mathbf{r}_\perp,\omega_j)}{\sqrt{N_j}}e^{i\beta_j z} ,\label{eq:EUnperturbed}\\
\mathbf{\tilde{H}}_{0}(\mathbf{r},\omega_j)&=& a_0\frac{\mathbf{h}_j(\mathbf{r}_\perp,\omega_j)}{\sqrt{N_j}}e^{i\beta_j z} . \label{eq:HUnperturbed}
\end{eqnarray}
where $\mathbf{e}(\mathbf{r}_\perp,\omega_j)$ and $\mathbf{h}(\mathbf{r}_\perp,\omega_j)$ are the vectorial electric and magnetic mode profiles. $a_0$ is the complex amplitude of the mode. $\beta_j$ is the mode propagation constant.  
The mode normalization and orthogonality condition read:
\begin{eqnarray}
\frac{1}{2} \int\{ \frac{\mathbf{e}_j(\mathbf{r}_\perp,\omega_j)}{\sqrt{N_j}} \times\frac{\mathbf{h}_i^*(\mathbf{r}_\perp,\omega_i)}{\sqrt{N_i}}\}\cdot \mathbf{\hat{z}} dA = \frac{1}{2} \int\{\frac{\mathbf{e}_i^*(\mathbf{r}_\perp,\omega_i)}{\sqrt{N_i}} \times\frac{\mathbf{h}_j(\mathbf{r}_\perp,\omega_j)}{\sqrt{N_j}}  \}\cdot \mathbf{\hat{z}} dA  = \delta_{ij}.\label{eq:normalization}
\end{eqnarray}
where the integral is performed on the plane perpendicular to the waveguide propagation direction and $\mathbf{\hat{z}}$ is the unit vector in the propagation direction $z$.
The normalization constant $N_j$ is set so that the total power propagating in the mode ($P_j$) equals $|a_0|^2$ : 
\begin{eqnarray}
P_j = \int\frac{1}{2}\Re \{|a_0|^2 \frac{\mathbf{e}_j(\mathbf{r}_\perp,\omega_j)}{\sqrt{N_j}}\times \frac{\mathbf{h}_j^*(\mathbf{r}_\perp,\omega_j)}{\sqrt{N_j}} \} \cdot \mathbf{\hat{z}} dA \equiv |a_0|^2 
\end{eqnarray}
We refer the reader to~\cite{snyder_optical_1983} for more information on the mode field distributions and their relation to one another. In this work the modes and the propagation constants are computed by use of a commercial mode solver (Lumerical). Yet we recall a couple of points that will play a role in the analysis of nonlinear coupling:
(i) The longitudinal electric field component of a mode has a phase difference of $\pi/2$ with the corresponding transverse components and (ii), because of the symmetry of the  index profile in the horizontal direction, the longitudinal and vertical electric field components display the opposite parity as the one of the horizontal component.
Here we consider airclad waveguides such that there is only one symmetry plane.
A few examples of the spatial distribution of the electric fields are shown in Fig.~\ref{figcrossing}.
Modes whose main component, the one with the strongest local electric field, is horizontal (resp. vertical) are labeled TE$_{lk}$ (resp. TM$_{lk}$), where $l$ and $k$ are the number of zeros in the horizontal and vertical directions~\cite{vermeulen_efficient_2010}.

\subsection{Nonlinear coupling}

We next derive the expressions for the nonlinear coupling between different forward propagating modes. 
In this derivation, the nonlinearity is treated as a perturbation to the ideal lossless linear waveguide.
The perturbed waveguide modes are written as: 
\begin{eqnarray}
\mathbf{\tilde{E}}(\mathbf{r},\omega_j)&=& a_j(z) \frac{\mathbf{e}_j(\mathbf{r}_\perp,\omega_j)}{\sqrt{N_j}}e^{i\beta_j z} ,\label{eq:EPerturbed}\\
\mathbf{\tilde{H}}(\mathbf{r},\omega_j)&=& a_j(z) \frac{\mathbf{h}_j(\mathbf{r}_\perp,\omega_j)}{\sqrt{N_j}}e^{i\beta_j z} .\label{eq:HPerturbed}
\end{eqnarray}
where $a_j(z)$ is the complex slowly varying amplitudes of the perturbed modes.
In practice, we consider the total field to be a superposition of a finite number of monochromatic waves:
\begin{eqnarray}
\mathbf{E}(\mathbf{r},t)&=& \sum_j\Re\{ a_j(z) \frac{\mathbf{e}_j(\mathbf{r}_\perp,\omega_j)}{\sqrt{N_j}}e^{i( \beta_j z-\omega_j t)}\} ,\\
\mathbf{H}(\mathbf{r},t)&=& \sum_j\Re\{ a_j(z) \frac{\mathbf{h}_j(\mathbf{r}_\perp,\omega_j)}{\sqrt{N_j}}e^{i( \beta_j z-\omega_j t)}\} .
\end{eqnarray}
These perturbed modes should obey the Maxwell curl equations including the nonlinear polarization $\mathbf{\tilde{P}}^{NL}(\omega_j)$,
\begin{eqnarray}
\nabla \times \mathbf{\tilde{E}}(\mathbf{r},\omega_j)&=& i\omega_j \mu_0 \mathbf{\tilde{H}}(\mathbf{r},\omega_j),\label{eq:MaxwellCurlEpert}\\
\nabla \times \mathbf{\tilde{H}}(\mathbf{r},\omega_j)&=& -i\omega_j \epsilon_0n^2 \mathbf{\tilde{E}}(\mathbf{r},\omega_j) -i\omega_j\mathbf{\tilde{P}}^{NL}(\mathbf{r},\omega_j).\label{eq:MaxwellCurlHpert}
\end{eqnarray}
To derive the coupled-wave equations, we start from the conjugated form of the Lorentz reciprocity theorem \cite{snyder_optical_1983}:
\begin{eqnarray}
\int\nabla \cdot \mathbf{F} dA = \frac{\partial}{\partial z} \int \mathbf{F} \cdot \mathbf{\hat{z}} dA .
\label{eq:divtheo}
\end{eqnarray}
The $\mathbf{F}$-field can be constructed from the perturbed and unperturbed waveguide mode fields as $ \mathbf{F} \equiv \mathbf{\tilde{E}}_{0}^*(\mathbf{r},\omega_j)\times \mathbf{\tilde{H}}(\mathbf{r},\omega_j) + \mathbf{\tilde{E}}(\mathbf{r},\omega_j)\times \mathbf{\tilde{H}}_{0}^*(\mathbf{r},\omega_j)$. Substituting this in Eq. \eqref{eq:divtheo} yields:
\begin{eqnarray}
\begin{aligned}
\int\{ (\nabla \times \mathbf{\tilde{E}}_{0}^*(\mathbf{r},\omega_j))\cdot \mathbf{\tilde{H}}(\mathbf{r},\omega_j) &- \mathbf{\tilde{E}}_{0}^*(\mathbf{r},\omega_j) \cdot (\nabla \times  \mathbf{\tilde{H}}(\mathbf{r},\omega_j)) \\ 
+ (\nabla \times \mathbf{\tilde{E}}(\mathbf{r},\omega_j))\cdot \mathbf{\tilde{H}}_{0}^*(\mathbf{r},\omega_j) &- \mathbf{\tilde{E}}(\mathbf{r},\omega_j) \cdot (\nabla \times  \mathbf{\tilde{H}}_{0}^*(\mathbf{r},\omega_j))\}  dA\\
&= \frac{\partial}{\partial z}\int\frac{a_0^* a_j(z)}{N_j} \{ \mathbf{e}_j(\mathbf{r}_\perp,\omega_j) \times\mathbf{h}_j^*(\mathbf{r}_\perp,\omega_j) + \mathbf{e}_j^*(\mathbf{r}_\perp,\omega_j) \times\mathbf{h}_j(\mathbf{r}_\perp,\omega_j))  \}\cdot \mathbf{\hat{z}} dA .
\end{aligned}
\label{eq:divtheo_subs}
\end{eqnarray}
The left hand side of Eq. \eqref{eq:divtheo_subs} can be simplified by substituting Eqs. \eqref{eq:MaxwellCurlE}-\eqref{eq:MaxwellCurlH} and \eqref{eq:MaxwellCurlEpert}-\eqref{eq:MaxwellCurlHpert} and the right hand side by using the normalization condition [Eq. \eqref{eq:normalization}]. This gives:
\begin{eqnarray}\label{eqamplitude}
\frac{{d}}{{d} z}a_j= i\omega_j\frac{e^{-i\beta_j z}}{4\sqrt{N_j}} \int\mathbf{e}_j^*(\mathbf{r}_\perp,\omega_j)\cdot\mathbf{\tilde{P}}^{NL}(\mathbf{r},\omega_j)dA.
\end{eqnarray}
\label{eq:divtheo_subs_final}

\section{Second harmonic generation}

We apply the formalism from the previous section to the specific case of second harmonic generation. For simplicity we only consider type I phase matching. The fundamental with a carrier frequency $\omega_0$ and the second harmonic with a carrier frequency $2\omega_0$ are each limited to a single spatial mode. The total electric and magnetic fields are:
\begin{eqnarray}\label{eqperturbedEfield}
\mathbf{E}(\mathbf{r},t)&=&\Re\{ a(z) \frac{\mathbf{e}_{a}(\omega_0, \mathbf{r}_\perp)}{\sqrt{N_a}}e^{i( \beta_a z-\omega_0 t)}+ b(z) \frac{\mathbf{e}_b(2\omega_0, \mathbf{r}_\perp)}{\sqrt{N_b}}e^{i( \beta_b z-2\omega_0 t)}\},\\
\mathbf{H}(\mathbf{r},t)&=&\Re\{ a(z) \frac{\mathbf{h}_a(\omega_0, \mathbf{r}_\perp)}{\sqrt{N_a}}e^{i( \beta_a z-\omega_0 t)} +b(z) \frac{\mathbf{h}_b(2\omega_0, \mathbf{r}_\perp)}{\sqrt{N_b}}e^{i( \beta_b z-2\omega_0 t)}\},
\end{eqnarray}
By injecting the fields in equation~\eqref{eqamplitude}, one finds the following coupled ordinary differential equation describing the nonlinear coupling between the two modes:
\begin{eqnarray}\label{eq_coupled_odea}
\frac{{d} a(z)}{{d} z}&=&i\omega_0\frac{e^{-i\beta_a z}}{4\sqrt{N_a}}\int{\mathbf{e}^*_a\cdot\mathbf{\tilde{P}}^{NL}(\mathbf{r},\omega_0)dA},\\
\frac{{d} b(z)}{{d} z}&=&i2\omega_0\frac{e^{-i\beta_b z}}{4\sqrt{N_b}}\int{\mathbf{e}^*_b\cdot\mathbf{\tilde{P}}^{NL}(\mathbf{r},2\omega_0)dA}.\label{eq_coupled_odeb}
\end{eqnarray}
In the time domain, the nonlinear polarization reads:
\begin{eqnarray}
\mathbf{P}^{NL}(\mathbf{r},t)=\Re\{\mathbf{\tilde{P}}^{NL}(\mathbf{r},\omega_0)e^{-i\omega_0t}+\mathbf{\tilde{P}}^{NL}(\mathbf{r},2\omega_0)e^{-i2\omega_0t}+\cdots\}
\end{eqnarray}
We here focus on a purely quadratic nonlinearity. By assuming a local response, we can write:
\begin{eqnarray}
\mathbf{P}^{NL}(\mathbf{r},t)=\varepsilon_0\iint\underline{\underline{\chi}}^{(2)}(\mathbf{r},t_1,t_2):\mathbf{E}(\mathbf{r},t-t_1)\mathbf{E}(\mathbf{r},t-t_2)dt_1dt_2.\label{eqnonlinearpolarization}
\end{eqnarray}
We now insert the electric field~\eqref{eqperturbedEfield} in \eqref{eqnonlinearpolarization} and find:
\begin{eqnarray}
\mathbf{\tilde{P}}^{NL}(\mathbf{r},\omega_0)&=&\frac{b(z)a^*(z)}{\sqrt{N_aN_b}}\varepsilon_0\underline{\underline{\chi}}^{(2)}(\mathbf{r},\omega_0;2\omega_0,-\omega_0):\mathbf{e}_b\mathbf{e}^*_ae^{i(\beta_b-\beta_a) z},\\
\mathbf{\tilde{P}}^{NL}(\mathbf{r},2\omega_0)&=&\frac{1}{2}\frac{a^2(z)}{N_a}\varepsilon_0\underline{\underline{\chi}}^{(2)}(\mathbf{r},2\omega_0;\omega_0,\omega_0):\mathbf{e}_a\mathbf{e}_ae^{i2\beta_a z},
\end{eqnarray}
where we introduced the commonly used Fourier components of the nonlinear tensor.
By injecting these expressions in~\eqref{eq_coupled_odea} and~\eqref{eq_coupled_odeb}, we find:
\begin{eqnarray}
\frac{{d} a(z)}{{d} z}&=&\frac{i\omega_0\varepsilon_0}{4}\frac{b(z)a^*(z)e^{i(\beta_b-2\beta_a) z}}{N_a\sqrt{N_b}}\int\sum_{jkl}\chi^{(2)}_{jkl}{e}^{*j}_a{e}^{k}_b{e}^{*l}_a  dA,\label{eq1}\\
\frac{{d} b(z)}{{d} z}&=&\frac{i\omega_0\varepsilon_0}{4}\frac{a^2(z)e^{i(2\beta_a-\beta_b) z}}{N_a\sqrt{N_b}}\int\sum_{jkl}\chi^{(2)}_{jkl}{e}^{*j}_b{e}^{k}_a{e}^{l}_a  dA,
\label{eq2}
\end{eqnarray}
where we expanded the tensor product $(j,k,l=x,y,z)$ and set $\underline{\underline{\chi}}^{(2)}(\mathbf{r},\omega_0;2\omega_0,-\omega_0) = \underline{\underline{\chi}}^{(2)}(\mathbf{r},2\omega_0;\omega_0,\omega_0) = \underline{\underline{\chi}}^{(2)}$. 

We define the effective nonlinear coefficient as:
\begin{eqnarray}\label{eqkappa}\\
\kappa = \frac{\omega_0\varepsilon_0}{4N_a\sqrt{N_b}}\int\sum_{jkl}\chi^{(2)}_{jkl}{e}^{*j}_b{e}^{k}_a{e}^{l}_a  dA,
\end{eqnarray}
 such that equations~\eqref{eq1} and~\eqref{eq2} become
 \begin{eqnarray}
\frac{{d} a(z)}{{d} z}&=&i\kappa^*b(z)a^*(z)e^{-i\Delta\beta z},\label{eq1kappa}\\
\frac{{d} b(z)}{{d} z}&=&i\kappa a^2(z)e^{i\Delta\beta z}.\label{eq2kappa}
\end{eqnarray}
where $\Delta\beta = 2\beta_a-\beta_b$.
In the literature, second harmonic generation is most often characterized by the undepleted theoretical conversion efficiency $P_{2\omega_0}(L)/(P_{\omega_0}(0)L)^2$ expressed in $\%$/(Wm$^2$) where $L$ is the length of the waveguide. By plugging the initial conditions $a(0)=\sqrt{P_{\omega_0}(0)},b(0)=0$ in equation~\eqref{eq2kappa}, and integrating over the length of the waveguide, neglecting pump depletion (as well as propagation loss), we obtain:
\begin{equation}\label{eqconveff}
|b(L)|^2 = |\kappa|^2|a(0)|^4L^2\mathrm{sinc}^2(\Delta\beta L/2).
\end{equation}
In the case of perfect phase matching ($\Delta\beta=0)$, equation~\eqref{eqconveff} becomes $P_{2\omega_0}(L)/(P_{\omega_0}(0)L)^2=|\kappa|^2$. In what follows we will use the theoretical conversion efficiency $|\kappa|^2$ to characterize second harmonic generation.

\section{Application to III-V-on-insulator wire waveguides}

We now focus on the specific case of III-V on insulator wire waveguides. 
Because the waveguide direction is not fixed in the crystal frame ($xyz$), we introduce new coordinates ($x'y'z'$) used to describe the optical wave in the waveguide frame. 
The propagation equations \eqref{eq1kappa} and \eqref{eq2kappa}, simply become
 \begin{eqnarray}
\frac{{d} a(z')}{{d} z'}&=&i\kappa^*b(z')a^*(z')e^{-i\Delta\beta z'},\label{eq1kappa}\\
\frac{{d} b(z')}{{d} z'}&=&i\kappa a^2(z')e^{i\Delta\beta z'}.\label{eq2kappa}
\end{eqnarray}
Most III-V wafers are grown along a crystallographic axis. Consequently, we may consider that the light propagates in the $xz$-plane (010) of the crystal. 
The two coordinate frames are linked through the rotation matrix:
\begin{eqnarray} 
\left(
\begin{array}{c}
x \\
y\\
z
\end{array}
\right)
=
\left(
\begin{array}{ccc}
\cos\theta & 0 & -\sin\theta \\
 0 & 1 & 0\\
\sin\theta & 0 & \cos\theta
\end{array}
\right)
\left(
\begin{array}{c}
x' \\
y' \\
z'
\end{array}
\right).
\end{eqnarray}
The zinc-blende crystalline arrangement of III-V crystals leads to a single nonzero tensor element ($\chi^{(2)}_{xyz} = 2d_{14} \neq 0$).
In that case the tensor product in~\eqref{eqkappa} becomes:
\begin{eqnarray}
\sum_{jkl}\chi^{(2)}_{jkl}{e}^{*j}_b{e}^{k}_a{e}^{l}_a  &=& \chi^{(2)}_{xyz}{e}^{*x}_b{e}^{y}_a{e}^{z}_a + 
\chi^{(2)}_{xzy}{e}^{*x}_b{e}^{z}_a{e}^{y}_a + \chi^{(2)}_{yxz}{e}^{*y}_b{e}^{x}_a{e}^{z}_a + \chi^{(2)}_{yzx}{e}^{*y}_b{e}^{z}_a{e}^{x}_a +  \chi^{(2)}_{zxy}{e}^{*z}_b{e}^{x}_a{e}^{y}_a + \chi^{(2)}_{zyx}{e}^{*z}_b{e}^{y}_a{e}^{x}_a \nonumber\\
&=& 2\chi^{(2)}_{xyz}\left({e}^{*x}_b{e}^{y}_a{e}^{z}_a+{e}^{*y}_b{e}^{x}_a{e}^{z}_a +{e}^{*z}_b{e}^{x}_a{e}^{y}_a\right)
\end{eqnarray}
where the second step is a consequence of the Kleinman symmetry condition~\cite{boyd_nonlinear_2003}.
The general form of the effective nonlinearity in a the crystal frame hence reads: 
\begin{eqnarray}
\kappa = \frac{\omega_0\varepsilon_0}{2N_a\sqrt{N_b}}\int\chi^{(2)}_{xyz}\left({e}^{*x}_b{e}^{y}_a{e}^{z}_a+{e}^{*y}_b{e}^{x}_a{e}^{z}_a +{e}^{*z}_b{e}^{x}_a{e}^{y}_a\right)  dA.
\end{eqnarray}
In the waveguide frame it becomes:
\begin{align}
\kappa = \frac{\omega_0\varepsilon_0}{2N_a\sqrt{N_b}}\int\chi^{(2)}_{xyz}&\left[\left({e}^{*x'}_b\cos\theta - {e}^{*z'}_b\sin\theta\right)\left({e}^{y'}_a\right)\left({e}^{x'}_a\sin\theta + {e}^{z'}_a\cos\theta\right)\right.\nonumber\\
&+\left({e}^{*y'}_b\right)\left({e}^{x'}_a\cos\theta - {e}^{z'}_a\sin\theta\right)\left({e}^{x'}_a\sin\theta + {e}^{z'}_a\cos\theta\right)\nonumber\\
&+\left.\left({e}^{*x'}_b\sin\theta + {e}^{*z'}_b\cos\theta\right)\left({e}^{y'}_a\right)\left({e}^{x'}_a\cos\theta - {e}^{z'}_a\sin\theta\right)\right]  dA. \label{eqkappafinal}
\end{align}
We now look for specific cases of phase matching to evaluate the theoretical conversion efficiency in III-V semiconductor waveguides.
We use indium gallium phosphide (InGaP) as the core material~\cite{dave_nonlinear_2015}. The $\chi^{(2)}_{xyz}$ coefficient was measured to be as high a 220~pm/V around 1550~nm~\cite{ueno_second_1997}. Because of the lack of birefringence in III-V semiconductors, several different approaches have been impemented to achieve phase matching, including form birefringence~\cite{fiore_phase_1998}, quasi-phase matching~\cite{eyres_all_2001} and modal phase matching~\cite{chang_heterogeneously_2018,ducci_continuous_2004}.
Here we rely on the latter. We use the material dispersion reported in~\cite{kato_optical_1994} to compute the effective index and spatial distributions of the optical modes.
We limit ourselves to second harmonic generation of a fundamental quasi-transverse electric mode (TE$_{00}$) and start with the dimensions of waveguides recently used for supercontinuum generation~\cite{dave_dispersive_2015}.
\begin{figure}[t]
\centering
 \includegraphics[width=16cm]{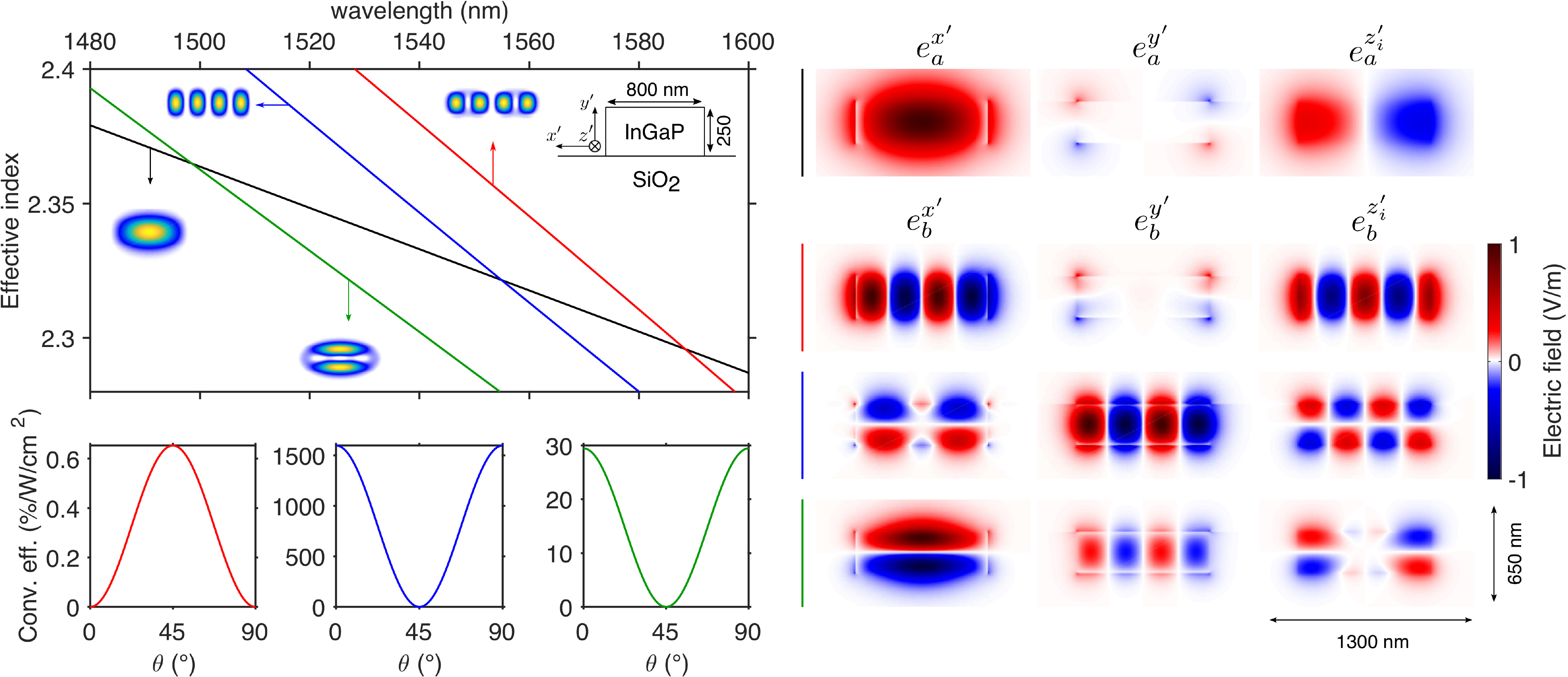}
\caption{Top left: Effective indices of a fundamental quasi-TE$_{00}$ pump mode and three different second harmonic higher order modes of a 250~nm high, 800~nm wide InGaP waveguide. The corresponding Poynting vector distributions are shown as inset. Bottom left: Theoretical conversion efficiency as a function of the propagation angle in the (010) crystal plane for the three phase matching points. Right: Spatial profiles of each component of the pump and SH modes.}
\label{figcrossing}
\end{figure}
Fig.~\ref{figcrossing} displays the index of the fundamental TE mode around the pump wavelength as well as several higher order modes around the second harmonic wavelength for a 800~nm wide, 250~nm thick InGaP on insulator wire waveguide. The corresponding spatial distributions of the electric fields are also shown. 
Note that the vectorial nature of the fields stands out in the figure as most components of the second harmonic modes have comparable magnitudes.
Several phase matching points, indicated by a crossing between the pump mode and a second harmonic mode, are found. The effective nonlinearity associated with each phase matching can be evaluated through equation~\eqref{eqkappafinal}. Its dependence with $\theta$ (also shown in Fig.~\ref{figcrossing}) reflects the $\bar{4}$ symmetry of the material. It can be leveraged for quasi-phase matching, as predicted in~\cite{dumeige_whispering_2006} and demonstrated in~\cite{kuo_second_2004}.
We here focus on straight waveguides and simply look for the angle that maximizes conversion between two modally phase matched waves. Interestingly it depends on the symmetry of the spatial distributions.
The conversion to the TE$_{30}$ mode is maximum when $\theta =\ang{45}$ while the conversion to the TM$_{30}$ and and TE$_{01}$ modes, is optimized when $\theta =\ang{0}$.  More generally, we only find maxima at $\ang{0}$ and $\ang{45}$ and hence focus on those two specific angles.

The $\ang{45}$ effective nonlinearity reads:
\begin{equation}\label{eqkappa45}
\kappa(\ang{45}) = \frac{\omega_0\varepsilon_0}{2N_a\sqrt{N_b}}\int\chi^{(2)}_{xyz}\left[{e}^{y'}_a\left({e}^{x'}_b{e}^{x'}_a-{e}^{z_i'}_b{e}^{z_i'}_a\right) + \frac{{e}^{y'}_b}{2}\left({e^{x'}_a}^2+{e^{z_i'}_a}^2\right)\right]  dA,
\end{equation}
where we introduced the spatial distribution $ e^{z_i'}=-ie^{z'}$ corresponding to the imaginary part of the longitudinal component.
To the best of our knowledge, previous results of second harmonic generation in III-V nanowaveguides were performed in this configuration~\cite{chang_heterogeneously_2018}. This is likely because the cleave directions for III-V semiconductors grown on (100) substrate are [110] and [1$\bar{1}$0]. Waveguides whose cleaved facets are perpendicular to the propagation direction are henced oriented $\ang{45}$ with respect to the crystal axis. For example, ultra efficient conversion was recently demonstrated in that direction in gallium arsenide wire waveguides~\cite{chang_heterogeneously_2018}, where a quasi-TE pump is coupled to a quasi-TM SH mode. In that case, it is the ${{e}^{y'}_b}{e^{x'}_a}^2$ term that dominates such that a scalar approximation suffices to predict the nonlinear coupling. Here however, we find conversion to a TE$_{30}$ mode whose vertical component is weak such that no single term dominates, highlighting the importance of a full vectorial approach even in a $\ang{45}$ waveguide.

The $\ang{0}$ effective nonlinearity, on the other hand, is:
\begin{equation}\label{eqkappa0}
\kappa(\ang{0}) = \frac{i\omega_0\varepsilon_0}{2N_a\sqrt{N_b}}\int\chi^{(2)}_{xyz}\left({e}^{x'}_b{e}^{y'}_a{e}^{z_i'}_a+{e}^{y'}_b{e}^{x'}_a{e}^{z_i'}_a -{e}^{z_i'}_b{e}^{x'}_a{e}^{y'}_a\right)  dA.
\end{equation}
Nonlinear coupling in this case always requires mixing between two different polarization components of the pump.
In both overlap integrals involving the TM$_{30}$ and TE$_{01}$ modes shown in Fig.~\ref{figcrossing}, it is the (${e}^{y'}_b{e}^{x'}_a{e}^{z_i'}_a$) term that dominates, as can be expected from using a quasi-TE pump.
Interestingly the $y'$ component of these second harmonic modes are similar despite the modes having very different Poynting vector distributions. The more efficient conversion is logically found when most of the energy propagates in the vertical electric field component.

The main difference between $\ang{0}$ and $\ang{45}$ oriented waveguides stems from the profiles of the excited modes. As a reminder there is a single, vertical, symmetry plane. Consequently, only the parity of the profiles along the $x'$ direction matters.
In a $\ang{45}$ waveguide, the $y'$ component of the SH mode must be symmetric as the fundamental components are squared [see Eq.~\eqref{eqkappa45}]. Conversely, in $\ang{0}$ waveguides, only SH modes with an antisymmetric vertical component will be excited because the product of the transverse and longitudinal components of the pump is always antisymmetric.
We stress that these considerations are valid for type I SHG, irrespective of the pump mode. 
Importantly we find that III-V nanowaveguides are suitable for efficient conversion in both propagation directions.
\begin{figure}[t]
\centering
 \includegraphics[width=16cm]{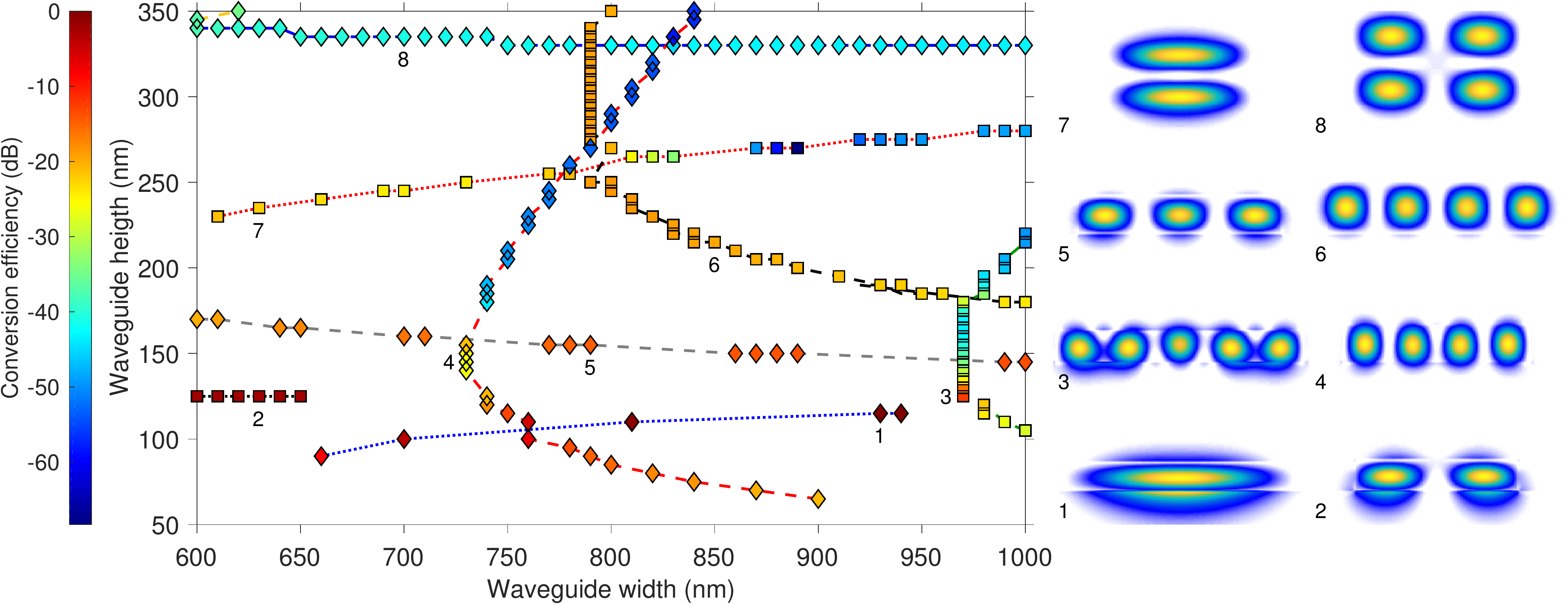}
\caption{Left: Efficiency map of the nonlinear coupling between a TE$_{00}$ pump mode and a higher order second harmonic mode. Only phase-matched interactions are shown. Diamond (resp. square) markers correspond to \ang{45} (resp. $\ang{0}$) waveguides. The lines connect neighboring points corresponding to the same higher order mode. Right: Poynting vector distribution of the eight independent second harmonic modes found in the map.}
\label{figmap}
\end{figure}

Next we study the impact of the waveguide dimensions when considering a pump around 1550~nm.
Specifically, we vary the width and height of the III-V section and look for crossings between a fundamental quasi-TE mode and a SH higher order mode in a 10~nm window around 1550~nm.
For each instance of phase matching, we compute the effective nonlinearity for different propagation directions and store only the maximum coefficient.  Due to the symmetry of the crystal we limit ourselves to the first quadrant. 
We investigate waveguides with a width between 600~nm and 1000~nm and a height between 50~nm and 350~nm. To limit the computational time, we used a resolution of 5~nm.
Every phase matching point is shown as a marker in Fig.~\ref{figmap}. The marker color codes the strength of the coupling and its shape indicates the angle between the waveguide and the crystal axis that maximizes the interaction. Squares are used for the $\ang{0}$ waveguides while diamonds represent $\ang{45}$ waveguides.
To highlight similar interactions, we evaluate the orthogonality between neighboring markers via equation~\eqref{eq:normalization}. We define a threshold at 70\%, beyond which we infer it is the same mode and connect the two markers with a line. We identify 8 independent modes. Their Poynting vector distribution is shown in Fig.~\ref{figmap}.
Unsurprisingly we find that the conversion increases with decreasing dimensions. 
The maximum effective nonlinearity [$\kappa = 2816\;\mathrm{(\sqrt{W}m)^{-1}}$] is found for a waveguide with a width of 810~nm and a height of 110~nm, directed at $\ang{45}$ degrees. The SH propagates in a TM$_{00}$ mode and the corresponding conversion efficiency is as high as 79300~\%/(Wcm$^2)$. This interaction is well known as it it mostly due to the mixing of the transverse components of the modes~\cite{chang_heterogeneously_2018}. 
More interesting are the many square markers indicating coupling enabled by longitudinal components. The maximum conversion efficiency [52350~\%/(Wcm$^2$)], found for  a TM$_{10}$ mode, is predicted to be almost as efficient as the conversion to a TM$_{00}$ mode. 
Moreover, in waveguides with a thickness of 200 nm or more, it is the $\ang{0}$ configuration that is the most efficient.
These are commonly used layers as the propagation is less impacted by inhomogeneities and surface roughness in that case, leading to lower propagation loss~\cite{ottaviano_low_2016,pu_efficient_2016}.

We stress that not all possible couplings are shown on the map. We use a 5~nm resolution and wave vectors can be very sensitive to waveguide dimensions. Also, quasi-phase matching can be used to efficiently couple two modes with different effective indices~\cite{parisi_algaas_2018}. Yet, many of the novel nonlinear couplings we show here are predicted to be very efficient. 
We expect them to play a significant role in future integrated wavelength converters.

\section{Conclusion}

We have theoretically investigated second harmonic generation in III-V semiconductor wire waveguides. By using a full vectorial model we found many instances of efficient conversion between a fundamental pump mode and a higher order second harmonic mode. Our results highlight the crucial role played by the longitudinal component of the electric field. When propagating along the crystal axis, only wave mixing involving different components is permitted by the single nondiagonal $\chi^{(2)}_{xyz}$ element. Due the high index contrast, the longitudinal electric field component can be almost as large as its transverse couterpart~\cite{driscoll_large_2009} making this configuration very efficient.

\section{Acknowledgements}

This work was supported by funding from the European Research Council (ERC) under the European Union’s Horizon 2020 research and innovation programme (grant agreement Nos 726420,  759483 \& 757800)  and by the Fonds de la Recherche Fondamentale Collective (grant agreement No PDR.T.0185.18).

\end{document}